
\normalbaselineskip=20pt
\baselineskip=20pt
\magnification=1200
\overfullrule=0pt
\hsize 16.0true cm
\vsize 22.0true cm
\def\lsim{\mathrel{\rlap{\lower4pt\hbox{\hskip1pt$\sim$}}
    \raise1pt\hbox{$<$}}}         
\def\gsim{\mathrel{\rlap{\lower4pt\hbox{\hskip1pt$\sim$}}
    \raise1pt\hbox{$>$}}}         

\def\overleftrightarrow#1{\vbox{\ialign{##\crcr
    $\leftrightarrow$\crcr
    \noalign{\kern 1pt\nointerlineskip}
    $\hfil\displaystyle{#1}\hfil$\crcr}}}
\long\def\caption#1#2{{\setbox1=\hbox{#1\quad}\hbox{\copy1%
\vtop{\advance\hsize by -\wd1 \noindent #2}}}}

\centerline{\bf Realistic Hadronic Matrix Element Approach to
Color Transparency }
\vskip 18pt
\centerline {B.K. Jennings}
\centerline {TRIUMF}
\centerline {Vancouver, B.C. V6T 2A3}
\vskip 18pt
\centerline{ G.A. Miller}
\centerline{Physics Department, FM-15}
\centerline{University of Washington, Seattle, Washington 98195}
\vskip 24pt
\centerline{\bf Abstract}
Color transparency occurs if a small-sized wave packet, formed in a high
momentum transfer process, escapes the nucleus before
expanding. The time required for the expansion
depends on the masses of the baryonic components of the wave packet.
Measured
proton diffractive dissociation and electron deep inelastic scattering
cross sections are used to
examine and severely
constrain the relevant masses. These constraints allow
significant color transparency effects to occur at experimentally accessible
momentum transfers.
\vskip 12pt
U.Wa. 1992 Preprint 40427-24-N92, TRIUMF preprint TR-pp-92-48,
hep-ph/9207211, Submitted to Phys. Rev. Lett.
\vfill \eject

\baselineskip=20pt
\vskip 12pt

Color Transparency (CT) is the postulated [1,2]
absence of final (or initial) state interactions
caused by the cancellation of color fields of a system of quarks and gluons
with small spatial separation.  For example, suppose an electron impinges on a
nucleus knocking out a proton at high momentum transfer.  The consequence of
color transparency is that there is no exponential loss of flux as the ejected
particle propagates through the nucleus.  Thus, the usually ``black" nucleus
becomes transparent. We restrict our attention to
processes for which the fundamental reaction is elastic, or at least a
two-body reaction.  This requires that the  nuclear excitation energy
be known well enough
to ensure that no extra pions are created.

  The existence of color transparency depends on: (1) forming a
small-sized wave packet in a high momentum transfer reaction. (2) the
interaction between such a small object and nucleons being suppressed (color
neutrality or screening) and (3) the wave packet escaping the nucleus while
still small.
That color neutrality (screening)
causes the cross section of small-sized color singlet
configurations with hadrons to be small was found in Ref. 3, and is
well-reviewed in
Refs. 4,5 and 6.  So we take item (2) as given. The others require more
discussion.

The formation of a small-sized wave packet (1) at
feasible energies is an open question even though asymptotic
perturbative QCD predicts that the size of
the ejected wave packet
is of order of the inverse of the momentum transfer Q. Including the
effects of gluon radiation  (Sudakov suppression) further increases
the importance of small separations between quarks [7] and,
as a consequence  leads to a faster
decrease with Q [8]. But, the minimum value of $Q$ required for the
wave packet to be small is not known.

It is also true that  at
experimentally available energies, the small object does expand as it moves
through the nucleus. Thus the final state interactions are suppressed but not
zero. The importance of this expansion was found by Farrar et al.~[9], and
by Jennings and Miller [10]. See also Ref. [11].

 Tantalizing but non-definitive evidence has been obtained in a pioneering
$(p,pp)$ experiment at {\it {Brookhaven National Laboratory}} (BNL) [12].
Color transparency is the object of current searches using electron and
proton beams [13,14].  The existence of color transparency has
not yet been demonstrated, and it would be useful to
improve the reliability of CT predictions.
Here we use
apparently unrelated diffractive dissociation (DD) and deep inelastic
scattering
(DIS) data to probe the existence of the small-sized wave packet and to
constrain the expansion process.

To be specific, consider the high $Q^2$ quasielastic $(e, e' p)$ reaction.
A wave packet is formed when a bound proton absorbs the virtual
photon. This wave packet is dubbed [4] a point like configuration $(PLC)$,
in an optimistic
notation.  Thus
$$|PLC \rangle = T_H (Q^2)|N\rangle, \eqno(1)$$
where the hard photon absorption operator is denoted as
$T_H (Q^2)$.
The $|N\rangle$ represents a nucleon at rest, and  $|N (\vec q) \rangle$
represents one of momentum $\vec q$.  The form factor is $F(Q^2)
= \langle N (\vec q) | T_H (Q^2) |N \rangle. $

We assume that the PLC has no soft interaction U with surrounding nucleons.
Then [6]
$$0 = U \, T_H \, (Q^2) | N \rangle . \eqno(2)$$
In the optical approximation
$U = - 4 \, \pi i \, Im \, \hat f \, \, \rho ,$
in which $\hat f$ represents the $PLC$ - nucleon
interaction as a sum of quark-nucleon
scattering operators and $\rho$ is the density of target nucleons.  Only the
dominant imaginary part of $\hat f$ is kept, and the nucleonic matrix element
$\langle N | 4 \pi \, Im \, \hat f \, | N \rangle = \sigma_p$, the
proton-nucleon total cross section.  Thus we may
abbreviate: $4 \pi
\, Im \, \hat f \equiv \hat \sigma$.  Taking the nucleon matrix element of
Eq.~(2) and using completeness yields
$$0 = \sigma_p + \sum_\alpha \, \int \limits_{(M + m_\pi)^2} d M^2_X \,
\langle N (\vec q) | \hat \sigma_p | \alpha, M^2_X \rangle \, {\langle \alpha,
M^2_X | T_H (Q^2) \, | N \rangle \over F (Q^2) } . \eqno(3a)$$
in which an intermediate state of mass $M^2_X$ has a set of quantum numbers
(including multiplicity) $\alpha$.  It is useful to define the integral term of
Eq.~(3a) as $I (Q^2)$.  Then
$$\sigma_p = - I \, (Q^2)\,. \eqno(3b)$$

As the $PLC$ propagates through a length $\ell$, each component acquires a
phase factor $e^{i \, p_X \ell}$ with $p_X^2 = p^2 + M^2_N - M_X^2$.  Then one
may define [15] an effective $PLC$-nucleon cross section, $\sigma_{eff}
(\ell)$:
$$\sigma_{eff} \, (\ell) \equiv \sigma_p \, + \, \int
\limits_{\alpha} \, \int \limits_{(M + m_\pi)^2}
d\, M^2_X \, \langle N (\vec q) | \hat \sigma | \alpha, M^2_X \rangle e^{i (p_X
- p) \ell}{\langle \alpha M^2_X | T_H(Q^2) \, | N \rangle \over F
(Q^2)}\eqno(4)$$

The reader may
wonder how Eq.~(3a) can ever be valid. This occurs in a model
obtained by Jennings and
Miller (JM)[10].  They represent the states $(\alpha \, M^2_X)$ by two-body
harmonic oscillator eigenfunctions $|N_m\rangle$
in two spatial dimensions ($|N_0\rangle\equiv |N\rangle$).  Then
\noindent
$T_H
(Q^2) | N \rangle \propto |\vec b = 0 \rangle$ and $\langle N_m | T_H \, (Q^2)
| N
\rangle = \langle N | T_H (Q^2) | N \rangle$=$F(Q^2)$.
Further JM take $\hat \sigma =
\sigma_p b^2/ \langle N | b^2 | N \rangle.$  Then $ \, b^2 | N \rangle =
\left[|N \rangle - | N_1 \rangle \right] / \langle N | b^2 | N \rangle$ where
$|N_1 \rangle$ is the symmetric $2 \hbar \omega$ state. This means that
$M^2_X = M^2_1$.
Using these relations in Eq.~(4)
gives $\sigma^{JM}_{eff}$
$$\sigma^{JM}_{eff} \, (\ell) = \sigma_p \left( 1 - e^{i \, (p_1 - p)
\ell}\right). \eqno(5)$$
We examine Eq.~(5) to  understand the results to be presented.
The quantity $(p - p_1)\approx (M_1^2-M_N^2)/2p$, which means that
${1\over (M_1-M_N)}\equiv \tau_0$ plays the role of a time scale for the
expansion of the  PLC.
If $\tau_0 << \ell$, the
two terms of eq(5) cancel and transparency occurs; otherwise, final state
interactions do occur. (The relevant value of $\ell$ is about
a nuclear radius.)

The previous two-state model has some desirable features, but it is not
 realistic because
 a continuum of proton states is excited in $pp\to pX$ reactions.
We therefore use experimental observations of the matrix elements appearing in
Eqs.~(3) and (4).  First we notice an apparent difficulty.  Those matrix
elements are off energy-shell extensions of scattering amplitudes.  The
violation of conservation of energy is approximately $\left(M^2_X -
M^2\right)/2p$, so that if the integrals converge for low values of $M^2_X$
(the virtuality is small) and
off shell effects can be neglected.  Then
$$\bigg|\langle N (\vec q) | \, \hat \sigma \, |\alpha, M^2_X \rangle\bigg|
\, = \,
\left[{d^2 \sigma^{DD} (\alpha) \over dt \, d M^2_X} \right]^{1/2} $$
$$\bigg|\langle \alpha, M^2_X | T_H \, (Q^2) | N \rangle \bigg| \, = \,
\left[{1 \over \sigma_M} \, {d^2 \, \sigma^{DIS} (\alpha) \over d \, \Omega
d \, E} \right]^{1/2} $$
where $DD$ and $DIS$ stand for diffractive dissociation and deep inelastic
scattering.  In $DD$ a fast proton breaks into the state $\alpha, M^2_X$
without exciting the bound target nucleon.  These are cross sections
in which the final state is denoted by the quantum numbers $\alpha$.
Define probabilities $P^{DD, DIS} (\alpha)$ so that
$$d \sigma^{DD, DIS} (\alpha) \, = P^{DD, DIS} (\alpha, M^2_x) \, d
\sigma^{DD, DIS},$$
where $\sum \limits_{\alpha} \, P ^{DD, DIS} (\alpha, M^2_X) = 1.$  This
is an often
used reasonable approximation.  Measurements of multiplicities [16,17] show
that $P^{DD, DIS} (\alpha)$ is a peaked but broad function of multiplicities.

One can see if existing data rule out 
Eq.~(3) by noting that
the integral term has a lower (negative) limit.  This can be obtained by
taking each product of matrix elements to be negative.  Then the quantity
$-I (Q^2)$ of Eq.~(3b) can be written as
$$\eqalignno{-I (Q^2) \, &\leq \, \int^\infty_{(M + M_\pi)^2} d \, M^2_X
\, \left[{d^2 \,
\sigma^{DD} \over dt \, d M^2_X} \, W_2 (x, Q^2) \right]^{1/2} \,
{\sum \limits_{\alpha}\left(P^{DD} (\alpha, M^2_X) \, P^{DIS} (\alpha,
M^2_X)\right)^{1/2} \over F (Q^2)} \cr
&\equiv I_{max}. &(6) \cr}$$
If $I_{max} < \sigma_p$, the data would rule out Eq.~(3).

We next evaluate $I_{max}$ to see if a $PLC$ can be formed.  We use Atwood's
[18] parameterization of $W_2 (x, Q^2)$ and Goulianos's [19] tabulation of
the s dependence of ${d^2 \sigma^{DD} \over dt d M^2_X}$
at t = -0.042 GeV$^2$
since much data are taken at that low value.
 The interpretation of the
$pp$ data is somewhat problematical, since the measurements
represent
the diffractive dissociation process only if ${M^2_X - M^2 \over s} \lsim
{m_\pi \over M}$[19], so that a maximum value of $M^2_X$ is given by $M^2_X \,
(max) \, \approx {m_\pi s \over M} + M^2$.  The probability functions
$P^{DD}, P^{DIS}$ are taken from Ref.~[16] for DIS and
Ref.~[17] for diffractive dissociation.  The sum over $\alpha$  is then
approximately 0.6,
approximately independent of $M^2_X$.  $I_{max}$ is evaluated by
performing the integral over $M^2_X$ up to a maximum value $M^2_c$.  If
$M^2_c$ exceeds $M^2_X (max)$ by a large amount, we would say $I_{max} <
\sigma_p$ and color transparency would be ruled out. The quantity $\sigma_p$ is
as tabulated in Ref. [20].

The use of the stated inputs shows that $I_{max}$ is greater than or equal to
$\sigma_p$ for values of $M^2_c$ between 2.4 and 2.6 GeV$^2$, depending
slightly
on s.  These values of $M^2_c$ have small virtuality and
do not exceed the bound required for
diffractive dissociation to occur.  Thus existing $DD$ and $DIS$ data allow the
existence of color transparency.  This is our strongest conclusion.

A further step is to use the above treatment of the integrand to
evaluate $\sigma_{eff}$ of Eq.~(4).
But  this could be  unrealistic: not all of the products of matrix
elements 
are negative and a sharp cutoff of the
$DD$  cross section is not expected.
In general, we should replace the factor
$\sum \limits_{\alpha} \, \left[P^{DD} (\alpha, M^2_X) \, P^{DIS} (\alpha,
M^2_X) \right]^{1/2}$ by the function $g (M^2_X)$:
$$g (M^2_X) = \sum_\alpha \, \left[ P^{DD} \, (\alpha, M^2_X) \, P^{DIS} \,
(\alpha, M^2_X) \right]^{1/2} \, {\rm {Sign}} \, (\alpha) \eqno(7)$$
where sign $(\alpha)$ is $\pm 1$ depending on the phases.  Measuring the
relative phases of $DD$ and $DIS$ amplitudes is difficult.  Thus although $g
(M^2_X)$ is a measurable function, it is not known.

It is reasonable to try a form
$g(M^2_X) = \left({M \over M_X}\right)^\beta $ (power-law) 
instead of the previously used
$g (M^2_X) = \theta (M^2_c - M^2_X) 0.6$ (sharp cutoff).~
 Values of $\beta$  ranging from 2.4 to 4.0 allow
the sum-rule relation (3)
to be satisfied at each value of $Q^2$.
The use of the power-law fall-off allows high mass $M^2_X$ states $(M^2_X
\approx Q^2)$ to participate in the integral without
emphasizing the importance of states of large virtuality.

We now turn to predicting nuclear color transparency. The
function $\sigma_{eff}$ is obtained by using the stated
inputs products of matrix elements.
The results are shown in
fig 1, for $s= 13 GeV^2$. (For electron scattering $s=Q^2+4 M^2$.)
If $g(M_X^2)$ is given by the power fall-off,
$\sigma_{eff}(\ell)\sim \ell$ for small
values of $\ell$. This similar to the model of
Ref [9]. If the sharp cut-off is used, one
obtains $\sigma_{eff}(\ell)\sim \ell^2$ for small
values of $\ell$.
$\sigma_{eff}$ is
generally smaller with the sharp cut-off
because
with $M_c^2\sim 2.2 GeV^2$
large values of $M_X$ do not appear, $p_X-p$ is prevented from
becoming large, and the cancellation between the
two terms of Eq.(4) is not disturbed much by the
phase factor $(p_X-p)\ell$.

Next we present predictions for the quasielastic
$(e,e',p)$ measurements being carried out
at SLAC [13]. The ratios of cross sections $\sigma/\sigma^{BORN}$ are shown
in Fig. 2.
The quantities $\sigma$ are (e,e'p) differential cross sections integrated
over the scattering angles of the outgoing proton. (See Ref.~10
for details.) Full color transparency
corresponds to a ratio of unity. We are concerned with the energies for
which $\sigma/\sigma^{BORN}$ approaches unity and for which it is
substantially greater than that obtained with the standard Glauber
treatment. Both choices of $g(M_X^2)$ show that
observable increases are
obtained for values of $\vec q$ as low as 5 GeV/c, or $Q^2=9$ GeV/c$^2$.
The results of using the sharp cutoff are very similar to those of using
the model of Ref. 10, with $M_1=1.44 GeV$. This follows from the small
value of $M_c$ and is also a consequence of the results shown in Fig.~1.

The single published experiment aimed at
observing the effects of color transparency  is
the BNL $(p,pp)$ work [12] at beam momenta
$p_L$ ranging from 6 to 12 GeV/c.
The kinematics of the BNL experiment are such
that the
basic $pp$ elastic scattering
occurs at
a center of mass angle of
90$^\circ$ if the target proton is at rest.
Fig.~3 shows
that the experimentally determined transmission $\sigma/\sigma^{BORN}$
 (ratio of nuclear to
hydrogen cross section per nucleon after removing the effects of
nucleon motion) has unexpected oscillations with energy.
Also shown is the expectation
 based on standard Glauber theory.
This standard  survived a rigorous examination
 in Ref. [21] and the independence on energy was
confirmed in a detailed
calculation that simulated the experimental conditions [22].

   One possibility,
 suggested by Ralston and
Pire [23], is that the energy dependence is caused by
an interference between a
hard amplitude which produces a small object, and a soft one (the
Landshoff process) which does not.
 Zakharov and Kopeliovich [11] and Jennings and Miller [24]
 have pursued this
by including the effects of the
expansion of the PLC. The technique is to use Eq.~(4) for the
initial or final state nuclear interactions of the small object and to use
the ordinary cross section $\sigma_p$ for those of the large object.
Another well-motivated mechanism is that of Brodsky and de Teramond [25]
in which
the  two-baryon system couples to charmed quarks (there is a
small (6q) and a large  (6q,$c\bar c$)
object) is also examined in Ref. [24].
None of these treatments reproduce the data.

Here we employ $\sigma_{eff}$ of Eq.~(4) as evaluated in Fig.~1.
 To approximate $T_H(Q^2)$ by $W_2^{1/2}$ is to assume that the
proton-proton high $Q^2$ data vary in a manner similar to $W_2$. This is
reasonable, because in each case  the reaction starts with a quark
absorbing high momentum. The Ralston-Pire mechanism is evaluated using a
recent more accurate fit of the hard pp scattering data by Carlson et al
[26]. Both the usual quark-counting and Landshoff amplitudes are included
in their description of $A_{nn}$ and the differential cross-sections.
 The results for the
mechanisms of Refs. [23] and [25] are
shown in Fig.~3. Both the power-law and sharp cutoff versions
of $g(M_X^2)$ are
used. We take these as representing lower and upper limits to the
predictions, and obtain a range of variation by shading the
area between these curves.
The
enhancement at about 4 GeV is a new consequence of the amplitude of
Ref.~[26]. The Brodsky-deTeramond model along with the sharp cutoff
$g(M_X^2)$ seems closest to the data. But
no calculation achieves good
agreement with the data.
One can say that the general trend is reproduced. The
strong dependence on $g(M_X^2)$ shows that at least one measurement of
color transparency is needed to determine this function.  The new
experiment [14] designed for higher energies and greater accuracy will
certainly help.

Our results are that measured
diffractive dissociation and deep inelastic scattering
data lend support to the idea that color transparency occurs.
The  formation of a PLC is allowed, and its expansion is not too rapid.
We eagerly await the new experimental results [13,14].
\vskip12pt
The authors acknowledge financial support from NSERC and USDOE, and thank
L.Frankfurt, W.R. Greenberg, and M. Strikman for useful discussions.

\vfill\eject
\noindent {\bf References and Footnotes}
\item{1.} A.H.~Mueller, ``Proceedings of Seventeenth Rencontre de
    Moriond", Moriond, 1982 ed. J Tran Thanh Van (Editions Frontieres,
Gif-sur-Yvette, France, 1982)p13.
\item{2.} S.J.~Brodsky in Proceedings of the Thirteenth
intl Symposium
on Multiparticle Dynamics, ed. W.~Kittel, W.~Metzger and A.~Stergiou (World
Scientific, Singapore 1982,) p963.
\item{3.}  S. Nussinov Phys. Rev.  Lett 34, 1286 (1975);
F.E. Low, Phys. Rev. D12, 163 (1975);
J Gunion D Soper Phys Rev. $ibid$ D15, 2617 (1977).
\item{4.} L. Frankfurt and M. Strikman, Progress in Particle and Nuclear
Physics, 27,135,(1991);
L. Frankurt and M. Strikman, Phys. Rep. 160, 235 (1988).
\item{5.} B.Z. Kopeliovich, Sov.J. Part. Nucl. 21, 117 (1990).
\item{6.} L Frankfurt, G.A. Miller \& M. Strikman, "Color Transparency and
Nuclear Phenomena", to be published Comm. Nuc. Part. Phys. Sept. 1992.
1992 UWA preprint-40427-26-N91.
\item{7.} H-n Li and G. Sterman,``The perturbative Pion Form Factor with
Sudakov Suppression", 1992 preprint ITP-SB-92-10.
\item{8.}L. Frankfurt, G.A. Miller \& M. Strikman, ``Precocious
Dominance of Point-like
configurations in Hadronic Form Factors", U.Wa. Preprint 40427-16-N92,
 submitted to Nucl.
Phys. A.
\item{9.}  G.R.~Farrar, H.~Liu, L.L.~Frankfurt \& M.I.~Strikman, Phys.
Rev. Lett. 61 (1988) 686.
\item{10.} B.K.~Jennings and G.A.~Miller, Phys. Lett. B236, 209 (1990);
 B.K.~Jennings and G.A.~Miller, Phys. Rev. D 44, 692 (1991);
 G.A. Miller and B.K. Jennings p. 25 in "Perspectives in Nuclear Physics
at Intermediate Energies" Ed. S. Boffi, C. Ciofi degli Atti, M. Giannini,
1992 ,World Sci. Press Singapore;
G.A.  Miller, ``Introduction to Color Transparency",
in ``Nucleon resonances and Nucleon Structure",  G.A.
Miller, editor. To be published by World Sci., Singapore (1992).
\item{11.}  B.Z.~Kopeliovich and B.G.~Zakharov, Phys. Lett. B264 (1991) 434.
\item {12.}A.S. Carroll et al Phys Rev Lett 61,1698 (1988; S. Heppelmann,
p. 199 in ``Nuclear physics on the Light Cone",ed. by M.B. Johnson and
L.S. Kisslinger , World Scientific (singapore, 1989).
\item {13.}SLAC Expt. NE-18, R. Milner, Spokesman.
\item {14.} A.S. Carroll {\it et al.}, BNL expt. 850.
\item{15.} The use of $\sigma_{eff}$ to replace the quantity $\sigma$ of
standard  scattering wave functions is a very  accurate approximation to a
more complete multiple scattering series.  See W. R. Greenberg and G.A.
Miller, ``Multiple Scattering Series for Color Transparency", U. WA.
1992 preprint 40427-23-N92, submitted to Phys. Rev.D.
\item{16.}See e.g. N. Schmitz p3, in ``Hadronic Multiparticle Production"
Ed. P. Carruthers, World Scientific, Singapore 1988.
\item{17.}K. Goulianos, H. Sticker and S.N. White PRL 48, 1454 (1982).
\item{18.}  A. Bodek {\it et. al.} Phys. Rev.D20,1471 (1979).
\item{19.} K. Goulianos, Phys. Rep. 101, 169 (1983)
\item {20.} PDG, Phys. Lett. B239,1 (1990)
\item {21.} T.-S.H. Lee and G.A.~Miller, Phys. Rev C. 45, 1863 (1992).
\item{22.} I. Mardor et al. "Effect of Multiple Scattering the measurement
of nuclear transparency" 1992 Tel Aviv Univ. preprint
\item {23.} J.P.~Ralston and B.~Pire, Phys. Rev. Lett. 61 (1988) 1823.
\item {24.} B. K. Jennings and G.A. Miller,
Phys. Lett. B 274,442 (1992).
\item {25.} S.J.~Brodsky \& G.F.~De~Teramond, Phys. Rev. Lett. 60
(1988) 1924.
\item {26.} C.E. Carlson ,M.Chachkhunashvili, and F. Myhrer ``Elastic
pp and $p\bar p\to \pi\pi$ reactions in short and middle distance QCD"
WM-92-101 , Feb. 1992

\vfill \eject
\noindent {\bf Figure Captions}
\bigskip
\item{1.}
The real part of $\sigma_{eff}(\ell)/\sigma$.
Dashed: sharp cut off $g(M_X^2)$,
dotted: eq.~(5) with   $M_1=1.44 GeV$,
dash-dot: power law $g(M_X^2)$.
\medskip
\item{2.} Ratios of cross sections for the $(e,e'p)$ reaction.
The solid line represents the standard Glauber calculation
($\sigma_{eff}=\sigma_p)$. The other curves are defined in Fig.~1.
\medskip
\item{3} Energy dependence of $\sigma/\sigma^{BORN}$.
Data points-Carroll et al.[12].
The area shaded vertically is obtained from the mechanism of Ref.~[23]
and amplitude of Ref.~[26].
 The area shaded horizontally
is obtained from the mechanism of Ref. ~[25].
Upper bound: sharp cut off $g(M_X^2)$
Lower bound: power law $g(M_X^2)$.
The solid curve assumes
no color transparency.

\medskip

\bye